\documentclass[aps,prb,twocolumn,showpacs]{revtex4}
\usepackage{amsmath}
\usepackage{subfigure}
\usepackage{graphicx}
\usepackage{epsfig}
\usepackage{txfonts}
\usepackage{epsf}

\newlength{\upit}\upit=0.1truein

\newcommand{\ltappr}{{{\lower4pt\hbox{$<$} } \atop \widetilde{ \ \ \ }}}
\newlength{\bxwidth}\bxwidth=1.5 truein

\newlength{\figwidth}
\newlength{\shift}
\usepackage{epsf}

\newcommand \bea {\begin{eqnarray} }
\newcommand \eea {\end{eqnarray}}

\newcommand{\beg}{\begin{equation}}
\newcommand{\en}{\end{equation}}

\newcommand{\dg}{^{\dagger }}
\newcommand{\bk}{\mathbf k}
\newcommand{\bq}{\mathbf q}

\newcommand{\br}{\mathbf r}

\begin{document}

\title{Correlated disorder in Kondo lattice}
\author{Maxim Dzero and Xinyi Huang}
\affiliation{Department of Physics, Kent State University, Kent, OH 44240, USA}

\date{\today}

\begin{abstract}
Motivated by recent experiments on Yb-doped CeCoIn$_5$, we study the effect of correlated 
disorder in Kondo lattice. Correlations between the impurities are considered at the two-particle level. 
We use mean-field theory approximation for the Anderson lattice model to calculate how the emergence 
of coherence in the Kondo lattice is impacted by correlations between impurities. We show that the 
rate at which disorder suppresses coherence temperature depends on the length of impurity correlations.
As impurity concentration increases, we generally find that the suppression of coherence temperature is significantly 
reduced. The results are discussed in the context of available experimental data. 
\end{abstract}


\maketitle

\section{Introduction}
Physical effects driven by an interplay of induced disorder and strong correlations in condensed matter systems 
have been one of the central topics of experimental and theoretical research for many years. One of the active research
directions is particularly focused on the properties of disordered heavy-fermion materials. In these materials, 
strong coupling between magnetic and non-magnetic degrees of freedom not only leads to a 
significant enhancement of the electron's mass below certain temperature $T^*$, but also to a number of remarkable
phenomena \cite{PiersReview}, such as unconventional superconductivity \cite{HFSC1,HFSC2,Cedo,PuCoGa5}, 
magnetic-field-induced non-Fermi liquid metallic phases \cite{YRS}, hidden-order phase transition \cite{URS} and, recently proposed
topological insulating phases \cite{TKI}. Studying how these systems react to disorder may help to better understand the microscopic mechanisms responsible for the formation of coherent states.

Earlier experiments have been mostly focused on the effect of doping on superconductivity and 
the onset of coherence (i.e. formation of heavy fermions) in Kondo lattices. 
It was generically found that a substitution of magnetic sites with non-magnetic impurities yields
a substantial reduction in both coherence and superconducting critical temperatures \cite{Steglich1983,exp1,exp2,Ott1987}. 
Later similar studies have been performed on materials belonging to the group of '115' heavy-fermion superconductors 
and have given qualitatively similar results \cite{Cigdem2010,Paglione} to what has been observed in other heavy fermions. 

The problem of how weak disorder affects the onset of coherence in Kondo lattice has 
been discussed theoretically  in Ref. [\onlinecite{Zlatko1986}]. It was assumed that impurity atoms
are independent and, therefore, the disorder correction to the conduction and $f$-electron self-energies were 
computed exactly within the self-consistent Born approximation. In particular, 
it was shown that impurities considerably affect the formation of the heavy fermion state when the strength 
of the disorder potential is comparable to the single site Kondo temperature. The dominant contribution to the suppression 
of $T^*$ comes from the disorder potential on $f$-sites.  It was pointed out in Ref. [\onlinecite{Zlatko1986}], this is not 
surprising given the fact that the heavy quasiparticle spectral weight is dominated by the $f$-states. Overall, the results of 
this theory \cite{Zlatko1986} were found to be in agreement with available experimental data. 

Recent measurements \cite{Cigdem2010,Maple2011,Fisk2011}, however, consistently demonstrated that the
substitution of Yb on Ce sites in CeCoIn$_5$ does not lead to significant changes in the properties of this material. 
Specifically, according to the data reported in Ref. [\onlinecite{Maple2011}], the onset of coherence remains essentially
unaffected with Yb-doping, while the critical temperature of the superconducting transition decreases linearly with 
Yb concentration.  Given the presence of strong correlations in this material, manifested by the
linear temperature dependence of resistivity at low temperatures \cite{NFL115} and unconventional $d$-wave superconducting pairing, these experimental results are highly unusual.

The recent experimental data motivate us to revisit the problem of disorder in heavy fermion materials. 
Specifically, in this paper we will address the following question: how the onset of coherence in the Kondo lattice as a function
of impurity concentration is modified by the correlations between the impurities? 
To answer this question, we will adopt the theory developed in Ref. [\onlinecite{Zlatko1986}] appropriately modified to include the correlations between impurities. Specifically, impurity correlations are described 
by the probability distribution function which depends on the distance between the two impurities.  
Within our theory we will show that even in the presence of weak disorder (i.e. when the self-energy corrections are small compared to the relevant energy scales in the system), the existence of the pair impurity correlations significantly slows the suppression of the coherence temperature. We also draw motivation from the experimentally observed enhancement of the superconducting critical temperature in high-$T_c$ superconductors caused by the ordering of the oxygen interstitials \cite{Oxigen2010}, as well as from a weak suppression of the superfluidity in $^3$He in the aerogel \cite{expHe3a,expHe3b,Fomin2008a,Fomin2008b}. 

The correlations between the impurities in a material can, in principle, be induced by the local lattice strains. 
To understand why the impurity correlations may slow down the suppression of the coherence temperature, let us first introduce the
characteristic length scale $R$ on which impurity distribution function significantly deviates from unity. Then, within the 
Born approximation, one can show that there will be two contributions to self-energy. One contribution, $\Sigma_{ii}$, corresponds to the scattering of electrons on the same impurity and, upon the averaging over disorder, this contribution is proportional to the concentration of impurities $n_{imp}$. The second contribution, $\Sigma_{ij}$, describes the scattering of electrons on two different impurities and, therefore, is proportional to $n_{imp}^2$. In the presence of impurity correlations, however, $\Sigma_{ij}$ becomes proportional to $n_{imp}^2R^3$. Thus if the radius of correlations is large enough (i.e. $n_{imp}R^3\sim 1$), 
$\Sigma_{ij}$ becomes comparable with the first, linear in $n_{imp}$, self-energy correction $\Sigma_{ii}$. 
Whether $\Sigma_{ij}$ can significantly compensate for $\Sigma_{ii}$ will, of course, 
depend on the specific form of the impurity distribution function, $p(r/R)$. For the specific form of $p(r/R)$, we will show that 
increase in $R$ yields higher values of $T^*$ compared to ones obtained for the case of uncorrelated disorder. 

Our paper is organized as follows. In Section II we discuss the possible origin of the impurity 
correlations in heavy-fermions and the self-energy corrections resulting from correlated disorder. In Section III we present
the calculation of the coherence temperature with the slave boson mean-field theory in the presence of correlated disorder. 
In Section IV we provide a general discussion of our findings in the context of available experimental data. Our results
are summarized in Section V. Finally, in Appendix A we discuss how the specific choice of the impurity correlation function
influences the electron scattering lifetime. 

\section{Preliminaries}
In this Section we define the impurity distribution form factor: the central quantity for our subsequent 
analysis of the coherence in the disordered Kondo lattice. In this Section for simplicity we limit the discussion 
to the problem of conduction electrons moving in the disorder potential, which is assumed to be weak. 

We will first demonstrate that the interactions between the impurities may be induced by the local strains in the host material. Then we proceed with evaluation of conduction electron self-energy, which will involve the impurity distribution form factor. 
Impurity distribution form factor formally appears due to averaging over the position of impurities. 
The form factor is, in turn, determined by the impurity correlation function.
We expect the impurity correlation function to be a functional of the interaction potential
between impurities, which allows us to approximately determine the shape of the former.  

\subsection{origin of the impurity correlations}
The two-impurity correlation can in principle be induced by the local lattice strains. 
To show this, let us first assume that the single impurity can be 
in only two states. The energy separating the these two states on impurity at site $i$ is $\Delta_i$. Then the impurities are described by the following
Hamiltonian:
\beg
H_{imp}=\sum\limits_i\Delta_i\hat{\sigma}_{iz}
\en
where ${\sigma}_{iz}$ is a third Pauli matrix and summation goes over the impurity sites. 

Generally speaking, the local ionic vibration of the impurity atom causes the change in $\Delta_i$. Expanding the energy splitting parameters up to the linear order in lattice displacements ${\mathbf u}_i$, we find
\beg
H_{imp}=\sum\limits_i\Delta_{i0}\hat{\sigma}_{iz}+\sum\limits_{\alpha\beta}\sum\limits_i\gamma_{\alpha\beta}^{(i)}\epsilon_{\alpha\beta}^{(i)}\hat{\sigma}_{iz},
\en
where $\Delta_{i0}$ is the value of the energy splitting at equilibrium, $\epsilon_{\alpha\beta}^{(i)}$ is a strain tensor at site $i$
\beg
\epsilon_{\alpha\beta}^{(i)}=\frac{1}{2}\left(\frac{\partial u_\alpha^{(i)}}{\partial x_\beta}+\frac{\partial u_\beta^{(i)}}{\partial x_\alpha}\right)
\en
and the components of $\gamma_{\alpha\beta}^{(i)}$ 
define the deformation potential. Consequently, the lattice deformation
at site $i$ may effect the impurity at site $j$. This can be shown by formally integrating out the fields $u_\alpha^{(i)}$ in 
the path-integral formulation of the problem. Alternatively one can use the condition for the balance between the internal and external stress in the system, which yields \cite{Halperin1977}:
\beg\label{Uij}
H_{imp}=\sum\limits_i\Delta_{i0}\hat{\sigma}_{iz}+\sum\limits_{ij}U_{ij}\hat{\sigma}_{iz}\hat{\sigma}_{jz}.
\en
The interaction potential decays as $U_{ij}\sim r_{ij}^{-3}$ at large distances. Most importantly, the sign of the interaction potential is defined by the local strain fields and can be either negative or positive, which corresponds to effective attraction or repulsion between impurities.

\subsection{disorder distribution: form factor}
To make our discussion self-contained, in this subsection we briefly review the theory of correlated disorder first discussed in Ref. [\onlinecite{Fomin2008a}] for the problem of the superfluid $^3$He in the porous medium. 
Hamiltonian describing the electrons moving in the disordered lattice reads:
\beg\label{Hc}
\begin{split}
H_c&=\sum\limits_\sigma\int d^3\br \psi_\sigma\dg(\br)\xi(\nabla)\psi_\sigma(\br)\\&+ u\sum\limits_\sigma\int d^3\br \sum\limits_i\delta(\br-{\mathbf R}_i)\psi_\sigma\dg(\br)\psi_\sigma(\br),
\end{split}
\en
where $u$ is a strength of the disorder potential, $\psi_\sigma\dg(\br)$ creation operator of an electron with spin projection $\sigma$ at point $\br$ in space, $\xi(\nabla)=-\nabla^2/2m-\mu$ and $\mu$ is a chemical potential.
For weak disorder, the second order correction to the electron's Green function 
${\cal G}(\br,\br';\tau)=-\langle\hat{T}_\tau \psi_\sigma(\br,\tau)\psi_\sigma(\br',0)\rangle$ in momentum and Matsubara frequency representation is \cite{AGD}
\beg
\begin{split}
&{\cal G}^{(2)}(\bk,\bk';i\omega)=|u|^2{\cal G}^{(0)}(\bk;i\omega){\cal G}^{(0)}(\bk';i\omega)\\ &\times
\sum\limits_{ij}\int\frac{d^3\bq}{(2\pi)^3}e^{-i(\bk-\bq){\mathbf R}_i-i(\bq-\bk'){\mathbf R}_j}{\cal G}^{(0)}(\bq;i\omega)
\end{split}
\en

To recover the spatially homogeneous expressions for the correlation functions, we follow the standard procedure \cite{AGD,Fomin2008a} of averaging over the impurity positions ${\mathbf R}_i$  :
\beg\label{ave}
\begin{split}
\sum\limits_i\langle e^{i(\bk-\bk'){\mathbf R}_i}\rangle_{dis}&=n_{imp}\delta(\bk-\bk'), \\ 
\sum\limits_{ij}\langle e^{-i(\bk-\bq){\mathbf R}_i-i(\bq-\bk'){\mathbf R}_j} \rangle_{dis}&=n_{imp}\delta(\bk-\bk')
[1+S(\bk-\bq)],
\end{split} 
\en
where we have introduced the impurity distribution form factor $S(\bk)$:
\beg
S(\bk-\bq)=n_{imp}\int d^3\br p(\br)e^{i(\bk-\bq)\cdot\br}
\en
Here $n_{imp}=N_{imp}/V$ is the concentration of impurities and $p(\br)$ is the impurity distribution function, which depends on the distance between two impurities. Thus, for conduction electron self-energy we obtain:
\beg\label{Sigma}
\Sigma(\bk,i\omega)=n_{imp}w_c^2\int\frac{d^3\bq}{(2\pi)^3}[1+S(\bk-\bq)]{\cal G}^{(0)}(\bq;i\omega)
\en

If we set the origin of the reference frame at one impurity, $p(\br)$ gives the probability to find a second impurity at distance $r$ from the first one. 
At large distances $p(r\to\infty)\to p_\infty$. After the proper choice of the normalization constant we can set $p_\infty=1$. Thus,
for uncorrelated imputiries $p(\br)=1$ and it follows $S(\bk)=n_{imp}\delta(\bk)$. In particular, this implies that the second term in the second equation (\ref{ave}) is of the order of $O(n_{imp}^2)$ and therefore can be neglected. 
However, as we will show below, for the correlated disorder, characterized by the correlation radius $R$, it is possible that the form factor becomes of the order $O(1)$ and thus it needs to be taken into account.

Since at large distances impurities are not correlated, expression for the form factor $S(\bk)$ contains both correlated and uncorrelated contributions. To single out the contribution from correlations we add and subtract the uncorrelated part:
\beg\label{Sk}
\begin{split}
S(\bk)&=n_{imp}\int d^3\br [p(\br)-1]e^{i\bk\cdot \br}+n_{imp}\delta(\bk)\\&\equiv
n_{imp}\int d^3\br v(\br)e^{i\bk\cdot \br}+n_{imp}\delta(\bk),
\end{split}
\en
where function $v(r)$ serves as the measure of impurity correlations. We can also omit the last term in (\ref{Sk}) since it yields a trivial correction of the order of $O(n_{imp}^2)$ to the self-energy. Therefore, to evaluate the momentum dependence of the form factor, we first will have to specify the function $p(\br)$ or, alternatively, correlation function $v(r)$.

From general considerations \cite{LLv5} it follows that the correlation function $v(r)$ should be a functional 
of the impurity interaction potential $U_{ij}=U{(|\br_i-\br_j|)}$, Eq. (\ref{Uij}). If we were to consider a weakly interacting 
ideal gas of atoms, for the pair of two atoms one could write the correlation function as 
$v_B(r_{ij})\sim [\exp(-U_{ij}/T)-1]$, where $T$ is a temperature (see Ref. \onlinecite{LLv5}). We will consider two correlation
functions of the following form \cite{Fomin2008a,Fomin2008b}:
\beg\label{vr}
\nu_\pm(r_{ij})=\pm\left[A_{\pm}\left(\frac{R}{r_{ij}}\right)^{3-\alpha}-1\right]e^{-r_{ij}/R},
\en
where $1<\alpha<3$, which also shows qualitatively similar dependence on distance as $v_{B}(r_{ij})$ (with corresponding attractive or repulsive potential $U_{ij}$). Our choice of the correlation functions (\ref{vr}) takes into account the possible fractal structure for the impurity atoms by means of the fractional exponent $\alpha$ and, as it turns out, makes the effects of the impurity correlations more pronounced compared to the ones described by $v_B(r_{ij})$. We allow for the possibility of having two types of correlation
functions (\ref{vr}), which qualitatively correspond to an effective attraction or repulsion between the impurity atoms on intermediate 
distances. This is done in order to understand better the origin of the "healing" effect in Kondo lattices, which we will discuss in the 
next Section. 

To evaluate the value of the constants $A\pm$ in Eq. (\ref{vr}), let us introduce the impurity density distribution function 
$n(\br)$. Consequently, the following correlation function is considered (for details, see discussion in Ch. XII of Ref. [\onlinecite{LLv5}]):
\beg\nonumber
\langle n(\br_1)n(\br_2)\rangle=n_{imp}\delta(\br_1-\br_2)+n_{imp}^2p(\br_1-\br_2).
\en
For the fluctuations in the impurity density one finds
\beg\label{densfl}
\langle \delta n(\br_1)\delta n(\br_2)\rangle=n_{imp}\delta(\br_1-\br_2)+n_{imp}^2v_{\pm}(r)
\en
Integrating both parts in Eq. (\ref{densfl}) with respect to the spacial coordinates and taking into account that in our case that the fluctuations in number of impurities are absent, we find \cite{Fomin2008a}
\beg\label{norm}
4\pi n_{imp}\int\limits_0^\infty v_\pm(r)r^2dr=-1.
\en 
The integral (\ref{norm}) with $v_\pm(r)$ given by (\ref{vr}) can be evaluated exactly. We find
\beg\nonumber
A_{\pm}=\frac{1}{\Gamma(\alpha)}\left(2\mp\frac{1}{4\pi n_{imp}R^3}\right).
\en
The momentum dependence of the form factor (\ref{Sk},\ref{vr}) can be evaluated exactly. We find
\beg\label{Sbk}
\begin{split}
S_\pm(k)=\pm{4\pi n_{imp}R^3}&\left(\frac{A_\pm\Gamma(\alpha-1)\sin[(\alpha-1)\tan^{-1}(kR)]}{kR[k^2R^2+1]^{(\alpha-1)/2}}\right.\\&\left.-\frac{2}{[1+(kR)^2]^2}\right)
\end{split}
\en
Note that the "strength" of the impurity correlations is effectively measured by the value of the dimensionless parameter
$n_{imp}R^3$.  From Eq. (\ref{Sbk}) we immediately observe that when the impurity correlation length scale $R$ is large enough, the form factor becomes of the order $O(1)$. In Appendix A we evaluate the effect of the impurity correlations on the electron's scattering
time, which is determined by the imaginary part of the self-energy. 

In the next Section we study how the correlated disorder will affect the onset
of the coherence on the Kondo lattice: a periodic lattice of predominantly localized $f$-electrons hybridized with 
conduction electrons.

\section{disordered Kondo lattice}
%
%
%
To describe the physics of an interplay between the disorder and coherence in the Kondo lattice we will 
employ the Anderson lattice model (ALM) in the limit of infinitely large Hubbard interaction $U$ between the localized $f$-electrons. 
Since the disorder is assumed to be weak, we first briefly discuss the
slave-boson mean-field approximation for the ALM and introduce the coherence temperature associated with the 
formation of the heavy fermion metallic state. We than proceed with the discussion of the disorder effects on the value of the coherence temperature for both correlated and uncorrelated disorder \cite{Zlatko1986}. 
\subsection{mean field theory of the Anderson lattice model}
We begin with writing down the model Hamiltonian to describe the physics of the disordered Kondo lattice. 
The Hamiltonian describing conduction electrons is 
\beg\label{Hcond}
H_c=\sum\limits_{\bk}\sum\limits_{\sigma=\uparrow,\downarrow}\xi_{\bk }c^{\dagger}_{\bk \sigma }c_{\bk \sigma}
\en
where $\xi_{\bk}=k^2/2m-\mu$ is the dispersion of
conduction electrons (assumed to be parabolic), $\mu$ is a chemical potential, 
$\sigma$ is a spin and $c_{\bk\sigma}^{\dagger}$ is a conduction electron creation operator. 
Consequently, the Hamiltonian which describes the $f$-electrons is:
\beg\label{Hf}
H_f=\sum\limits_{j,\alpha=1}^{N_{\Gamma}}\epsilon_{f} f_{j\alpha}\dg f_{j\alpha}+
{U}\sum\limits_{i\alpha\alpha'}f_{i\alpha}\dg f_{i\alpha}f_{i\alpha'}\dg f_{i\alpha'}.
\en
where $f_{j\alpha}\dg$ creates an $f$-electron on site $j$ in a state 
$\alpha$ of a lowest lying multiplet $N_\Gamma$-degenerate multiplet of the $f$-ion, 
$\epsilon_{f}$ is the $f$-electron energy and $U>0$ is the strength of the Hubbard interaction between the $f$-electrons. We note that index $\alpha$ is not a spin index due to the presence of the strong spin-orbit coupling.  Generally states belonging to the multiplet $\Gamma$ are described by the total angular momentum $J$ and $z$-component $M$ or some linear superposition of those states and in the second term (\ref{Hf}) the summation is restricted to $\alpha\not=\alpha'$.

Finally the term describing how conduction electrons are hybridized with localized $f$-electrons is
\beg\label{Hh}
H_h=\sum\limits_{j,\alpha=1}^{N_\Gamma}\left[V {c}_{i\alpha}^{\dagger} {f}_{j\alpha}+ V^*{f}_{j\alpha}\dg {c}_{i\alpha}\right],
\en
Here $V$ is a hybridization matrix element between the conduction electrons and localized $f$-electrons and $c_{i\alpha}\dg$
are project the conduction states with spin $\sigma$ onto an $f$-state classified by the 
$z$-component of angular momentum $M$. This projection is furnished by the presence of strong spin-orbit coupling 
\cite{CoqblinSchrieffer} and, generally speaking, also yields the hybridization matrix element to be non-local. However,  
as we have verified, the realistic spatial (or momentum) dependence of the hybridization elements  does not 
affect our main results and, therefore, will be ignored here. Thus, in the absence of disorder the periodic 
Anderson model Hamiltonian reads:
\beg\label{HPALM}
H_{PAM}=H_c+H_f+H_h
\en

In heavy-fermion materials Hubbard interaction is much larger than the bandwidth, so that we can take it formally to infinity,
$U\to\infty$. In this case, the multiply occupied states, $f^2$, $f^3$ etc., on which $H_{PAM}$ (\ref{HPALM}) operates must be projected out \cite{Barnes}. This is achieved by introducing the slave boson field $b_i$, which act like projectors 
\cite{SlaveBoson1,SlaveBoson2,SlaveBoson3,MillisLee}. As a result for our model Hamiltonian we can write:
\beg\label{Hpamprojected}
\begin{split}
H_{PAM}=&H_c+\sum\limits_{j,\alpha=1}^{N_\Gamma}\left[V {c}_{j\alpha}^{\dagger} b_j\dg{f}_{j\alpha}+ V^*{f}_{j\alpha}\dg {c}_{j\alpha}b_j\right]\\ &+\sum\limits_{j,\alpha=1}^{N_{\Gamma}}\epsilon_{f} f_{j\alpha}\dg f_{j\alpha}.
\end{split}
\en
It has to supplemented by the constraint condition of not allowing for more than one $f$-electron per site:
\beg\label{constraint}
Q_i=\sum\limits_{\alpha=1}^{N_\Gamma}f_{i\alpha}\dg f_{i\alpha}+b_i\dg b_i=1.
\en

To find the ground state governed by the model (\ref{Hpamprojected},\ref{constraint}) the mean-field approximation is adopted. 
Within the mean-field theory, the slave-boson operators  are replaced with their average values:
\beg\label{sba}
b_i\to \langle b_i\rangle\equiv a,
\en
which is then needs to be determined self-consistently. The self-consistency equations are derived from minimizing the
free energy with respect to the slave boson amplitude $a$ and also Lagrange multiplier $\lambda$ used to enforce the constraint
(\ref{constraint}). In addition, the total number of particles needs to be conserved. As a 
result one obtains the following system of three nonlinear equations \cite{SlaveBoson3,Barzykin2006}:
\beg\label{mfeq}
\begin{split}
(\lambda-\epsilon_f)a+VT\sum\limits_{\bk,\omega_n}{\cal G}_{fc}(\bk,\omega_n) &= 0, \\
(a^2-q_\Gamma)+T\sum\limits_{\bk,\omega_n}{\cal G}_{ff}(\bk,\omega_n)&= 0, \\ 
(q_\Gamma-a^2)+T\sum\limits_{\bk,\omega_n}{\cal G}_{cc}(\bk,\omega_n)=C,
\end{split}
\en
where $T$ is a temperature, $q_\Gamma=1/N_\Gamma$ and $C$ is a constant (for heavy fermion metals $C\simeq 1/2$). 
Lastly, the propagators entering into Eqs. (\ref{mfeq}) are given by:
\beg\label{propagators}
\begin{split}
{\cal G}_{cc}(\bk,\omega_n)&=\frac{i\omega_n-\epsilon_f}{(i\omega_n-\epsilon_f)(i\omega_n-\xi_\bk)-V^2a^2}, \\
{\cal G}_{fc}(\bk,\omega_n)&=\frac{Va}{(i\omega_n-\epsilon_f)(i\omega_n-\xi_\bk)-V^2a^2}, \\
{\cal G}_{ff}(\bk,\omega_n)&=\frac{i\omega_n-\xi_\bk}{(i\omega_n-\epsilon_f)(i\omega_n-\xi_\bk)-V^2a^2}, \\
\end{split}
\en
The mean-field approximation becomes exact in the limit $N_\Gamma\to\infty$ while for the finite values of $N_\Gamma$ the corrections to mean-field results are of the order of $O(1/N_\Gamma)$.  We note, however, that the slave-boson 
mean-field theory proved to be very reasonable approximation for many heavy-fermion materials \cite{Hewson,PiersReview,Barzykin2006}. 

The nonzero solution for the amplitude $a$ appear for the first time at temperature
\beg
T_0^*\simeq D\exp\left[-\frac{|\epsilon_f|}{\rho_FV^2}\right],
\en
where $D$ is a width of the conduction band and $\rho_F$ is a density of states at the Fermi level. Non-zero value of $a$ signals the opening of the hybridization gap in the single particle spectrum. Opening of the gap also yields the enhancement of the electron's mass  \cite{MillisLee}. On the other hand, experimental data shows that in heavy fermion metals resistivity has a maximum at some temperature, which is usually associated with $T_0^*$, and then decreases upon further cooling \cite{PiersReview,Cedo,TransportCeCoIn5b}. Therefore, $T_0^*$ is interpreted as a coherence temperature below which the heavy-fermion metallic state is formed. 

\subsection{effect of disorder on coherence in Kondo lattice}
In this Section we discuss the effect of the correlated disorder on the coherence temperature $T^*$. Although we are
well aware of the fact, that slave-boson mean-field theory may not be entirely reliable approximation to compute $T^*$, 
we believe that it suffice to demonstrate disorder correlations affect the value of $T^*$ in comparison with 
uncorrelated case.

Since disorder breaks translational symmetry of the lattice, the slave-boson 
amplitude (\ref{sba}) as well as other mean-field parameters become inhomogeneous. 
However, for the case of weak disorder, i.e. when the self-energy corrections to the conduction and $f$-electron 
propagators are assumed to be much smaller than $T^*$, we can perform disorder averaging. This would yield the parameters
of the theory homogeneous, so that to determine the value of $T^*$ in the presence of disorder we still need to solve
the system of equations (\ref{mfeq}) with propagators including the self-energy corrections due to disorder \cite{Zlatko1986}.   

The disorder is simulated by the following Hamiltonian:
\beg\label{Hdis}
H_{dis}=\sum\limits_{j,\alpha}\left[W_{fj}f_{j\alpha}\dg f_{j\alpha}+W_{cj}c_{j\alpha}\dg c_{j\alpha}+
(W_{mj}f_{j\alpha}\dg b_jc_{j\alpha}+\textrm{h.c.})\right]
\en
We assume that random variables $W_{fi}$, $W_{ci}$ and $W_{mi}$ 
are Gaussian random variables, satisfying $\langle W_{f,c,m;i}\rangle_{dis}=0$
and $\langle W_{f,c,m;i}W_{f,c,m;j}\rangle_{dis}=W_{f,c,m}^2\delta_{ij}$, where $\langle{...}\rangle_{dis}$ 
denotes the averaging over the random distribution of $W_{f,c,m}$'s \cite{Zlatko1986}. 

Disorder corrections to self-energy are most compactly written using the matrix notation for the electron propagators:
\beg
\hat{\cal G}_0(\bk,i\omega)=\left[
\begin{matrix}
{\cal G}_{ff}(\bk,i\omega) & {\cal G}_{fc}(\bk,i\omega) \\
{\cal G}_{cf}(\bk,i\omega) & {\cal G}_{cc}(\bk,i\omega)
\end{matrix}
\right].
\en
Within the Born approximation for the self-energy we find:
\beg\label{SelfEnergy}
\hat{\Sigma}(\bk,i\omega)=n_{imp}\int\frac{d^3\bq}{(2\pi)^3}{\mathbf W}\hat{\cal G}_0(\bq,i\omega){\mathbf W}
\left[1+S_{+}(\bk-\bq)\right],
\en
where the disorder averaging has already been carried out, impurity correlation function $S_{+}(k)$ is given by (\ref{Sbk}) with $\alpha=3/2$ and the elements of the matrix ${\mathbf W}$ are
self-evident. Thus, the propagators entering into Eqs. (\ref{mfeq}) should be replaced with
\beg\label{GExact}
\hat{G}^{-1}=\hat{\cal G}_0^{-1}-\hat{\Sigma}.
\en
In principle, we could have used the self-consistent Born approximation (SCBA) which amount to replacing the bare propagator in (\ref{SelfEnergy}) with an exact one (\ref{GExact}) and solving for the self-energy self-consistently. This task can be easily accomplished 
for the uncorrelated disorder when the self-energy does not depend on momentum. In the presence of the disorder correlations, 
however, we have chosen the ladder approximation for the self-energy in order to facilitate the numerical calculations. We have also
verified, that the self-energy computed within SCBA does not differ significantly from the ones given by (\ref{SelfEnergy}) for 
a wide range of the disorder strength. In addition, to keep our  results tractable we will adopt the same approximation as in Ref. [\onlinecite{Zlatko1986}] and ignore the disorder potential in hybridization, $W_m=0$. We have checked, that this approximation does not lead to any qualitative changes in the values of $a$ and $T^*$ for both correlated and uncorrelated disorder.   

\begin{figure}[h]
\includegraphics[width=3.3in,angle=0]{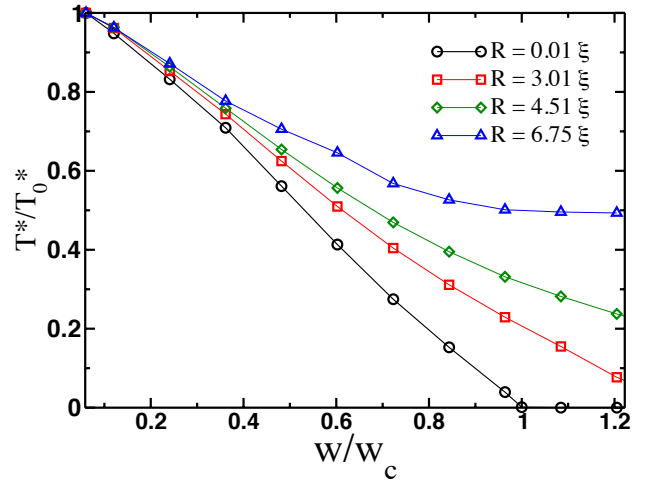}
\caption{Dependence of the coherence temperature (in the units of the coherence temperature for the clean system) on impurity
concentration $n_{imp}$ multiplied by the square of the disorder potential, $w=n_{imp} W^2$. Normalization parameter $w_c$ corresponds to the critical value of disorder when the coherence temperature vanishes in the absence of impurity correlations.}
\label{FigT}
\end{figure}

We evaluate the coherence temperature $T^*$ for fixed disorder potential 
$W=W_f=W_c=0.85 T_0^*$. This specific choice for disorder strength is dictated by the fact that both slave-boson amplitudes and
coherence temperature should be significantly suppressed when disorder becomes of the order of the single ion Kondo temperature \cite{Zlatko1986}. This, in turn, should make an effect of disorder correlations more pronounced. 
On Fig. \ref{FigT} we present our results of the numerical solution of Eqs. (\ref{mfeq}) for $T^*/T_{0}^*$ as a function of disorder $w=n_{imp}W^2$ for various values of the impurity correlation radius $R/\xi$ ($\xi=v_F/T_0^*$).  

From Fig. \ref{FigT} we see that impurity correlations have small effect at small concentrations, which is expected given the fact the 
self-energy corrections due to scattering off two impurities is proportional to $n_{imp}R^3\ll 1$. As $n_{imp}R^3$ increases, the suppression of the coherence temperature becomes less pronounced. It is also important to remember that as the value of the 
parameter $n_{imp}R^3$ reaches a certain critical value, the ladder approximation, used to evaluate the correlation functions, 
breaks down. This is the reason why in Fig. 1 we have restricted the range of disorder parameter variations to the value
$w/w_c=1.25$. We also find that the the same calculation with the correlation function $\nu_{-}(r)$, Eq. (\ref{vr}), yields even stronger suppression of $T^*$ as correlation radius increases, so that no "healing" effect occurs.

\section{Discussion}
%
%
%
The rate at which the coherence temperature is suppressed as the correlation radius is changing clearly depends on 
the specifics of the spacial dependence of the impurity correlation 
function $\nu_{\pm}(r)$, Eq. (\ref{vr}). The correlation function $\nu_{+}(r)$
at distances $r\geq R$ is negative which can be interpreted as an effective attraction between the impurity atoms. Thus 
we lead to interpret the observed enhancement (in comparison with uncorrelated case) of $T^*$ as being due to 
the cluster formation of Yb atoms. In this sense our interpretation is therefore similar to the one discussed in Ref. [\onlinecite{Fisk2011}]. The difference, however, is that the electronic state, although being spatially inhomogeneous due to disorder, 
is not strongly coupled to the impurities and the reduction in the $T^*$ suppression rate is a property of an impurity system
itself: there is no special re-organization of the electronic system in response to disorder. 

In addition to insensitivity of $T^*$ to Yb-doping in CeCoIn$_5$, equally puzzling property of this material is an unexpectedly slow
suppression of the superconducting transition temperature $T_c$ \cite{Maple2011,Fisk2011}. One intriguing question in this context
is whether concentration dependences of both $T^*$ and $T_c$ are correlated with each other. 
We believe that this question can certainly 
be addressed within the recently proposed theory of the tandem pairing in CeCoIn$_5$ \cite{Flint2010}. In this theory, 
the Cooper pairing emerges as a result of strong hybridization between singly occupied $f$-states and conduction electrons in 
the presence of two conduction channels emerging from the fluctuations into empty and doubly occupied $f$-states. The symmetry
of the conduction channels is determined by the point group irreducible representations and, as a result, ultimately determines
the symmetry of the superconducting order parameter. Within the large-$N$ mean field theory, however, the emergence of the
superconductivity corresponds to the opening of the second hybridization gap at the Fermi level. Therefore, within this picture, 
the behavior of both $T^*$ and $T_c$ will have to be correlated. We are planning to address this question 
in the future.

Another important question is how the presence of correlations between the impurity atoms can be probed experimentally. 
We think that one way to probe impurity correlations is to use the scanning tunneling microscopy (STM) measurements. Using this technique, one should be able to measure the position of impurities and any non-trivial physics associated with them. 
Alternatively, one can use momentum space probes such as X-ray diffraction spectroscopy \cite{Oxigen2010} to resolve for the presence of any fractal structures in the array of Yb atoms.

\section{Conclusions}
In this paper we have revisited a problem of how disorder affects the formation of  heavy-fermions in Kondo lattice. Specifically, 
we have studied how the presence of correlations between the impurity atoms influences the value of the coherence temperature. 
Our calculations have been performed within the framework of the slave-boson mean-field theory. Although this theory may not
provide an entirely reliable estimate for the coherence temperature, it still allows us to study the effects of impurity correlations. 
In addition, we have also restricted ourselves to the ladder approximation to evaluate the disorder corrections to self-energy. 
This approximation limits the range of applicability of our theory to the case of strongly correlated impurities, 
but still is reliable when the correlations are not very strong. 
In agreement with previous studies we found that in the absence of correlations the coherence temperature is strongly suppressed with increase in impurity concentration or values of disorder potential. When the correlation radius becomes comparable to the heavy-fermion coherence length $\xi$, we find the suppression of the coherence temperature is significantly reduced. 
Our theory can be directly applied to the case of Yb-doped CeCoIn$_5$ to account for the dependence of coherence temperature
on Yb concentration. In particular, we suggest that the Yb impurities are strongly correlated in this material, which leads to insensitivity 
of coherence temperature to disorder. The specific type of correlation potential can, in principle, be probed by the STM measurements.

\section{acknowledgments}
We would like to thank E. Abrahams, C. Almasan, P. Coleman, G. Kotliar, and K. Quader for discussions. 
Authors acknowledge the financial support by the Ohio Board of Regents Research Incentive Program grant OBR-RIP-220573.
This work was supported in part by the National Science Foundation under grant No. 1066293 and the hospitality of the Aspen Center for Physics (M.D.) 

\begin{appendix}
\section{electron scattering time due to correlated impurities}
\begin{figure}[h]
\includegraphics[width=3.2in,angle=0]{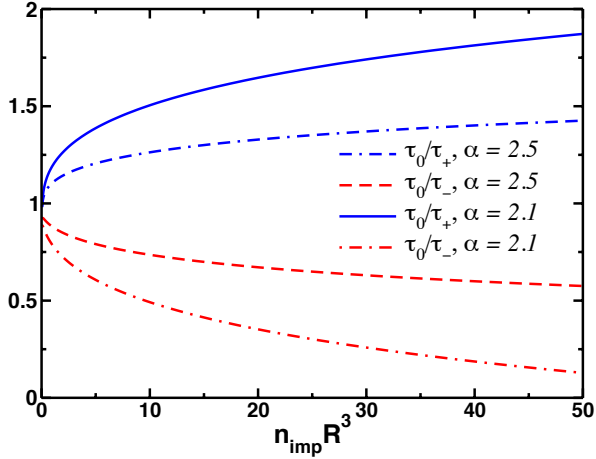}
\caption{Plot of the electron's relaxation times $\tau_{\pm}$, Eq. (\ref{taupm}), 
for correlated weak disorder as a function of the dimensionless parameter
$n_{imp}R^3$ for two values of the exponent $\alpha$, Eq. (\ref{vr}). 
We use the collision time $\tau_{0}$ for uncorrelated disorder to set the scale. 
As we have expected, the correlations
between the impurities lead to an increase between the electron impurity scattering events.}
\label{Fig1}
\end{figure}

In this Section we evaluate the changes in the time between two impurity scattering events due to the correlated disorder.  
We define the electron's scattering time due to disorder as follows
\beg\label{taupm}
\frac{\hbar}{\tau_{\pm}}=\text{Im}\left[\Sigma_\pm(k_F,0)\right],
\en
where $p_F$ is a Fermi momentum. We evaluate $\tau_{\pm}$ numerically using (\ref{Sigma}) and impurity correlation functions
(\ref{vr}) at $T\to 0$ and fixed impurity concentration $n_{imp}=1 \%$ (per unit volume) as a function of the dimensionless parameter $n_{imp}R^3$. We show the results calculation on Fig \ref{Fig1}. As we can see from Fig. \ref{Fig1}, depending on the sign of the correlation function $\nu_{\pm}(r)$, electron's scattering time is either drastically increased or discreased with an increase in impurity correlations until impurity correlation parameter $g\sim 1$. For higher degree of correlations, the changes in scattering time are moderate.

Our results in this Section demonstrate how the presence of correlations between impurities affect the scattering properties of conduction electrons. By comparing the signs of the functions $v_B(r_{ij})$ and $v_{\pm}(r_{ij})$, Eq. (\ref{vr}), 
we can associate the sign of the impurity interaction potentials with the sign of the
correlation function. In particular, it follows that 
the interaction described by the correlation functions $\nu_{+}(r)$ are repulsive at short distances $r\ll R$ and attractive at
large distances $r\gg R$, interactions $\nu_{-}(r)$ are repulsive for all distances. Our results can be qualitatively understood as follows: the repulsion (attraction) between impurities means a larger (smaller) mean-free path for conduction electrons and, as a consequence, increase (decrease) in scattering time. 
Independent of the sign of the form factor, however, the effect of impurity correlations is the renormalization of the
total scattering cross section. 

\end{appendix}

\end{document}